\documentclass[11pt,a4paper]{article}

\usepackage{scientificdata-template}
\usepackage{comment}
\usepackage{graphicx}
\usepackage{subcaption}
\usepackage{xurl}

\newcommand{\SciSchema}{\href{https://scischema.org/}{SciSchema.org}}
\newcommand{\SchemaMiner}{\href{https://github.com/sciknoworg/schema-miner}{\textsc{schema-miner}}}
\newcommand{\modelref}[2]{\href{#1}{#2}}

\title{\SciSchema{}: A Multidisciplinary Collection of Schemas for Structured Scientific Process Descriptions}

\author[1,$\dagger$]{Jennifer D'Souza\thanks{The first six authors are listed in order of contribution. The first three formed the founding team: Jennifer D'Souza led the project and performed end-to-end schema-mining experiments for 15 processes; Sameer Sadruddin led template publication and maintenance; and Anisa Rula contributed to conceptualization and coordination across all phases. Authors four onward contributed through domain expertise, schema design, data curation, and validation of their respective schemas; authors four to six additionally conducted the end-to-end experiments for their contributed process. The remaining contributor groups are ordered alphabetically by the last name of the first-listed member of each group, with internal group order preserved, and contributed equally.}}
\author[1]{Sameer Sadruddin}
\author[2]{Anisa Rula}
\author[3]{Ana Bossler}
\author[3]{Andr\'es Fullana}
\author[3]{Enric Bas}

\author[4]{Syed Ather}
\author[5]{Defne Circi}
\author[5]{Anlan Chen}
\author[5]{L. Catherine Brinson}
\author[6]{Alyssa Columbus}
\author[7]{George Demetriou}
\author[8]{Dongjun Jeong}
\author[9]{Tarun Kumar}
\author[10]{Frank Kr\"uger}
\author[10]{Sascha Genehr}
\author[11]{Kai Budde-Sagert}
\author[12]{Anamaria Leonescu}
\author[13]{Francesco Lodola}
\author[13]{Chiara Florindi}
\author[14]{Gagana Balasubramanya Murthy}
\author[15]{Samson Oluwapelumi Olagbile}
\author[16]{Nazia Riasat}
\author[17]{Yan Sha}
\author[18]{Kevin Shen}
\author[17]{Shaokai Yang}

\affil[1]{TIB Leibniz Information Centre for Science and Technology, Hannover, Germany}
\affil[2]{University of Brescia, Brescia, Italy}
\affil[3]{University of Alicante, Alicante, Spain}
\affil[4]{Georgia Institute of Technology, Atlanta, United States}
\affil[5]{Duke University, Durham, United States}
\affil[6]{Johns Hopkins University, Baltimore, United States}
\affil[7]{University of Manchester, Manchester, United Kingdom}
\affil[8]{Narnia Labs, Daejeon, South Korea}
\affil[9]{Hewlett Packard Enterprise Labs, India}
\affil[10]{Wismar University of Applied Sciences, Wismar, Germany}
\affil[11]{University of Rostock, Rostock, Germany}
\affil[12]{University College London, London, United Kingdom}
\affil[13]{University of Milano-Bicocca, Milan, Italy}
\affil[14]{Cambridge Institute of Technology, Bengaluru, India}
\affil[15]{Dangote Fertiliser Limited, Lagos, Nigeria}
\affil[16]{North Dakota State University, Fargo, United States}
\affil[17]{University of Alberta, Edmonton, Canada}
\affil[18]{SES AI, Woburn, United States}
\affil[$\dagger$]{Correspondence: jennifer.dsouza@tib.eu}
\date{}

\begin{document}

\maketitle

\begin{abstract}
Scientific processes are often described in heterogeneous article discourse, with details needed for comparison, reproducibility, reuse, and automation dispersed across prose, tables, figures, protocols, and supplementary files. We present the first release of \SciSchema{}, a multidisciplinary collection of 16 expert-annotated schemas spanning Biology \& Biotechnology, Materials \& Chemistry, Imaging \& Measurement, Physics, and Psychology. Each schema defines reusable fields for describing process instances, including inputs, outputs, materials, instruments or software, parameters, conditions, procedural steps, measurements, and provenance-related information. The schemas were created through a human-in-the-loop schema-mining workflow in which large language models generated candidate structures from process specifications, scientific articles, and expert feedback, followed by domain-expert construction of final master schemas. The dataset contains final schemas in JSON Schema and SHACL formats, intermediate model-generated schemas, expert-feedback records, source-paper metadata, community-development materials, and analysis scripts. Technical validation assessed schema structure, development provenance, expert review, and syntactic conformance. The collection supports structured annotation, metadata enrichment, scientific knowledge graphs, information extraction, semantic publishing, and cross-study comparison.
\end{abstract}


%https://www.nature.com/sdata/submission-guidelines
%Data Descriptors: describe new, open research datasets in a manner that promotes reuse, without reporting whether datasets support hypotheses or conclusions. They contain details of how datasets were created (Methods), what they contain (Data Records), and how they were checked and validated (Technical Validation), alongside any code used to create them. 

\section*{Background \& Summary}

%%%%%%%% WHAT %%%%%%%%%%%
Scientific research is judged by both its outputs and how they were produced, yet the information needed to understand, compare, reproduce, and reuse it often remains embedded in prose, figures, tables, supplementary files, or local laboratory conventions rather than structured descriptions. Scholarly and open-science infrastructures---including Dataverse \cite{magazine2011dataverse}, DataCite \cite{brase2009datacite}, Crossref \cite{hendricks2020crossref}, CORE \cite{knoth2023core}, OpenAlex \cite{priem2022openalex}, OpenCitations \cite{peroni2020opencitations}, OpenAIRE \cite{manghi2012openaireplus}, the European Open Science Cloud \cite{ayris2016realising}, arXiv \cite{ginsparg2011arxiv}, and the Open Research Knowledge Graph (ORKG) \cite{auer2025open}---support publication, citation, storage, interlinking, organization, and discovery, but do not close this process-level metadata gap. Here, a \emph{scientific process} is an experimental, computational, or analytical activity class that transforms inputs into outputs under specified materials, instruments, software, parameters, conditions, and constraints; its metadata characterize execution and support interpretation, comparison, reproducibility, and reuse. \SciSchema{} is a community-maintained collection of FAIR \cite{wilkinson2016fair} process schemas defining which attributes to record, how to organize them, and how to represent them in machine-actionable form. The schemas capture the relevant ``what, how, with what, under which conditions, and with what result'' \cite{ghiringhelli2023shared} through inputs, outputs, materials, instruments, parameters, settings, conditions, procedural steps, and provenance. They are released as JSON Schema for document- and software-oriented reuse and SHACL/Notation3 for RDF and knowledge-graph reuse, with both representations encoding the same content. They define process classes and fields for describing reported instances, not populated article-level records of individual executions. \SciSchema{} thereby makes scientific methods more transparent, comparable, interoperable, and reusable. Its initial release comprises 16 schemas spanning biology and biotechnology, materials science and chemistry, imaging and measurement, physics, and psychology.

%%%%%%%% WHY - 2 paragraphs %%%%%%%%%%%

%interdisciplinarity in science -- a bottleneck for lack of knowledge sharing infrastructures 
%the current paradigm of document-based scientific publishing no longer serves ... central to the ORKG
% several problems: reproducibility crisis, deficiency of peer review, predatory publishing, lack of interdisciplinarity
%how can scischemas.org facilitate interdisciplinarity? by discovering knowledge structures as schemas for disciplines, it is easier to establish links between what would have otherwise been discourse silos.

\begin{figure}[!tb]
    \centering

    \begin{subfigure}{\textwidth}
        \centering
        \includegraphics[width=\textwidth]{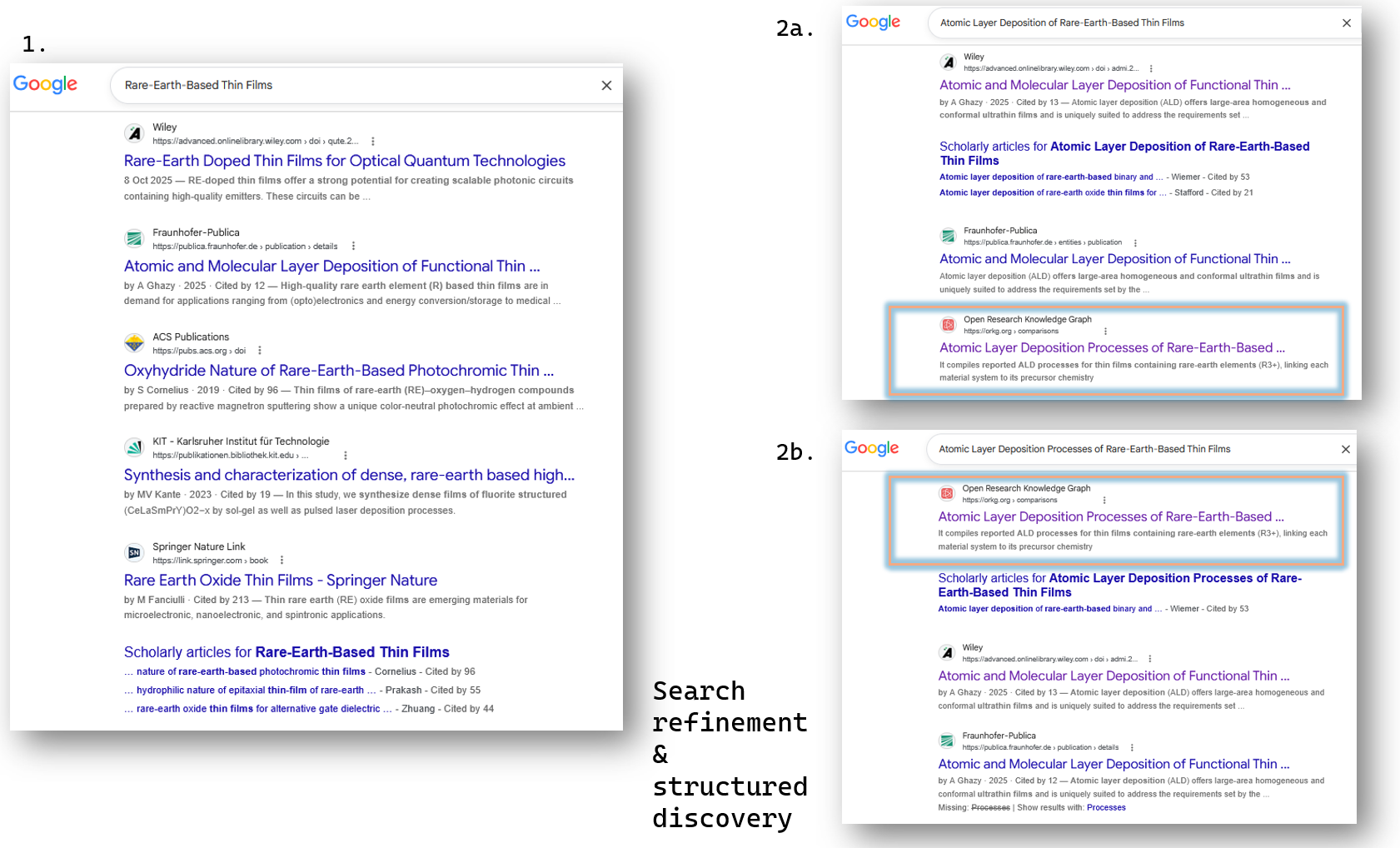}
        %\caption{}
        %\label{fig:search-to-structured-insight-a}
    \end{subfigure}

    %\vspace{0.5em}

    \begin{subfigure}{\textwidth}
        \centering
        \includegraphics[width=\textwidth]{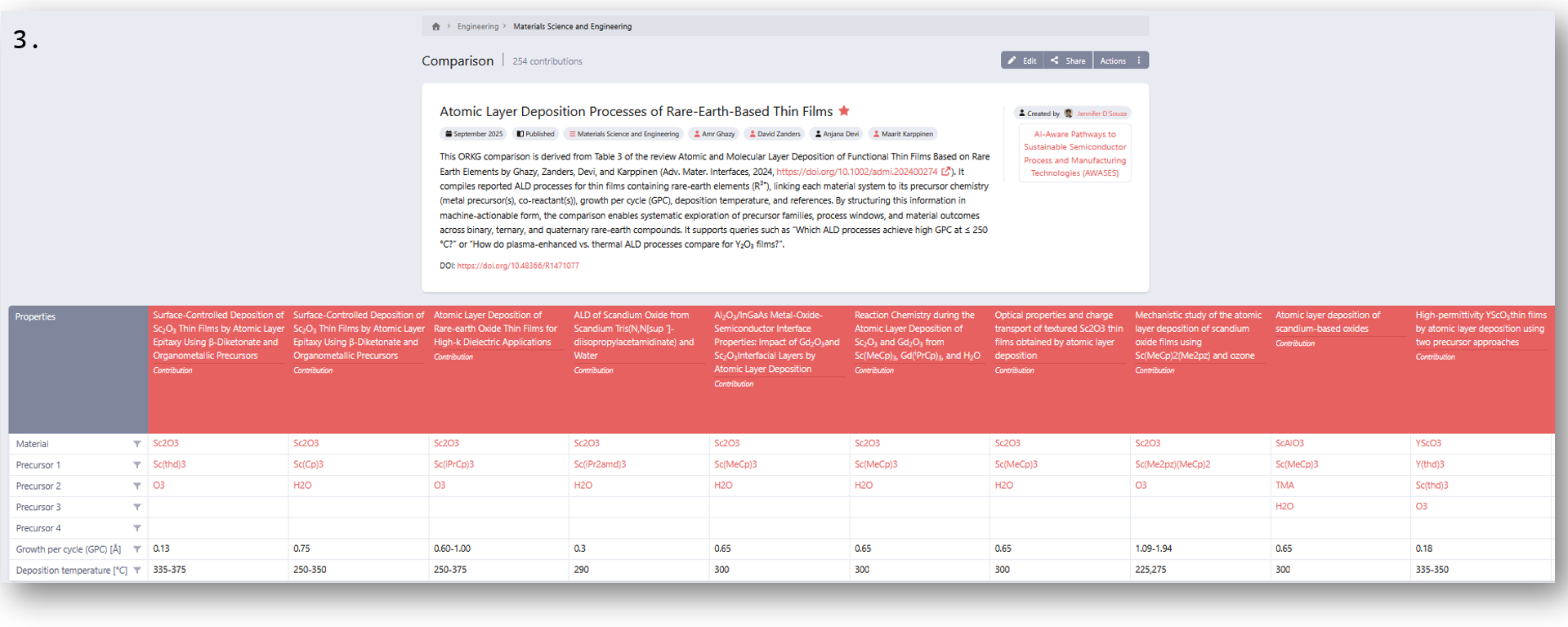}
        %\caption{}
        %\label{fig:search-to-structured-insight-b}
    \end{subfigure}

\caption{\textbf{\textit{Panel} 1.} A broad query returns mostly full-text scholarly articles in search results. \textbf{\textit{Panel} 2a.} A more method-specific query surfaces a schema-based structured comparison among the top five results. \textbf{\textit{Panel} 2b.} A more process-specific query surfaces the same structured comparison at the top of the search results. \textbf{\textit{Panel} 3.} Snapshot of the full structured comparison \url{https://orkg.org/comparisons/R1471077}, based on a simple illustrative schema for Atomic Layer Deposition processes of rare-earth-based thin films. The comparison includes more than 200 research papers represented through reported values for seven schema properties. Although this Open Research Knowledge Graph (ORKG) \cite{auer2020improving} example uses a simple schema, it illustrates the broader \SciSchema{} motivation: fine-grained process schemas can make method- and process-level details more findable and comparable, reducing the cognitive burden of retrieving them from prose.}
    \label{fig:search-to-structured-insight}
\end{figure}

Current scholarly communication remains predominantly document-centric: search engines retrieve publication pages, peer review evaluates manuscripts, and readers must reconstruct scientific content from prose, figures, tables, and supplementary files. This creates a cognitive bottleneck because the process information needed to compare studies---including materials, instruments, parameters, conditions, measurements, and outputs---is not exposed as a structured layer over the article. Incomplete and inconsistent methodological reporting further hinders interpretation, reuse, and replication: in a \emph{Nature} survey of 1,576 researchers, more than 70\% reported having failed to reproduce another scientist's experiment, while an analysis of COVID-19 randomized controlled trials found persistent deficiencies in reporting transparency, completeness, and consistency \cite{baker20161,kapp2022transparency}. Schemas cannot ensure complete reporting or reproducibility, but they define common process attributes that make reported information explicit, comparable, and queryable and, when records distinguish unreported from inapplicable information, help reveal reporting gaps. As illustrated in Figure 1, this enables a shift from document-centric retrieval to structured process discovery: method- and process-specific queries can surface an ORKG comparison in which papers are organized by reported process attributes \cite{auer2020improving}. Although the example uses only seven properties, it shows how standardized attributes can become findable, comparable, and traceable rather than reconstructed manually from prose. \SciSchema{} provides this schema layer by defining, for each process class, the attributes through which studies can be described and compared, including inputs, materials, instruments, parameters, conditions, procedural steps, outputs, and provenance-relevant descriptors. The schemas are not article-level process records; they define how such records can be instantiated, queried, compared, and linked to scholarly sources, thereby supporting systematic, traceable, and machine-actionable representations of scientific methods.

%disciplinary fragmentation → literature fragmentation → AI/automation pressure → need for process schemas → policy convergence → final claim
The need for \SciSchema{} is heightened by increasingly interdisciplinary, automated, and AI-mediated science. Major research problems require knowledge to cross disciplinary boundaries, although institutions, funding, and evaluation have historically remained discipline-based \cite{metzger1999interdisciplinary}. This fragmentation is mirrored in differing vocabularies and reporting conventions, even where processes share common structures of inputs, procedures, conditions, measurements, and outputs. Large-scale efforts such as the Human Genome Project \cite{international2001initial}, AlphaFold \cite{jumper2021highly,tunyasuvunakool2021highly}, the Large Hadron Collider \cite{atlas2026higgs}, and LIGO \cite{abbott2016observation} show that modern discovery relies on coordinated pipelines of data generation, instrumentation, computation, modeling, and interpretation. Meanwhile, AI-for-science, large language models (LLMs), agentic systems, and coding assistants are reshaping research search, synthesis, planning, and execution \cite{eger2025transforming,zhang2025exploring,zheng2025automation}, while automated laboratories require explicit representations of experimental intent, parameters, instruments, workflows, and outputs \cite{hillson2019building,mehr2020universal,boiko2023autonomous,tom2024self,canty2025science}. Prose-based methods are therefore inadequate as a common interface across disciplines, infrastructures, and computational agents. \SciSchema{} addresses this gap by preserving domain-specific detail while defining shared, community-curated dimensions for cross-process comparison. This role aligns with open-science and data-policy frameworks that treat interoperability as a condition for reuse, including the FAIR principles \cite{wilkinson2016fair}, UNESCO's Open Science Recommendation \cite{unesco2023openscience}, and the OECD recommendation on publicly funded research data \cite{oecd2021researchdata}. Similar priorities appear in U.S. public-access and data-sharing policies \cite{nelson2022ostppublicaccess,nih2023dmspolicyoverview}, China's scientific-data measures \cite{statecouncil2018scientificdata}, Japan's FAIR-aligned policies \cite{integratedinnovationstrategy2024openaccess,jst2025openaccessdatamanagement}, Canadian and Australian data-stewardship policies \cite{triagency2021rdmpolicy,nhmrc2025opensciencepolicy}, and European initiatives on the European Open Science Cloud (EOSC), common data spaces, data governance and access, and public-sector interoperability \cite{europeancommission2026datastrategy,europeanunion2022datagovernanceact,europeanunion2023dataact,europeanunion2024interoperableeuropeact}. Scientific process schemas are therefore more than documentation aids: they provide a structured layer for reproducible scholarship, cross-domain comparison, AI-ready knowledge graphs (KGs), and machine-assisted or automated experimentation in which scientific intent, evidence, and action must be computable.

This positioning complements related efforts in structured scientific metadata, workflow representation, and linked-data publication. \href{https://schema.org/}{Schema.org} provides a Web-scale vocabulary for embedding structured data in Web pages and improving machine interpretation \cite{guha2016schema-org-paper}; for scientific data, it has mainly supported dataset discovery, Web markup, and high-level machine-interpretable resource descriptions \cite{gray2023schemaorg}. \href{https://bioschemas.org/}{Bioschemas} extends this approach to the life sciences through community profiles and usage guidelines over Schema.org types, improving the findability and interoperability of biological resources \cite{gray2017bioschemas,bioschemas}. Other initiatives operate at different layers: the Chemical Description Language (\(\chi\)DL) represents executable organic-chemistry procedures \cite{mehr2020universal}, MaiML provides an execution-centric interchange format for laboratory measurement records \cite{nishio2025digital}, and nanopublications support provenance-aware publication of small scientific assertions as linked data \cite{groth2010anatomy,kuhn2018nanopublications}. \SciSchema{} addresses an adjacent need by providing community-curated, process-level schemas that transform prose-bound scientific procedures into comparable, machine-actionable metadata models across domains.

%%%%%%%% HOW %%%%%%%%%%%
%DEFINE schema mining
Constructing scientific-process schemas requires agreement on which entities and attributes to represent, how to organize domain terminology, and how to maintain clarity, coverage, and reuse as practices evolve. A schema defines the structure for describing process instances, whereas an ontology represents domain concepts and their semantic relationships. Although \SciSchema{} does not provide ontologies, ontology engineering offers a relevant precedent for the consensus, quality control, and sustained community maintenance required by shared knowledge resources; the Gene Ontology exemplifies this model \cite{ashburner2000gene,neuhaus2022ontology,matentzoglu2022ontology}. Scientific-process schema construction is especially challenging because relevant properties are dispersed across prose, tables, figures, protocols, and supplementary materials, while assessing completeness requires domain expertise. We define \emph{scientific schema mining} as a human-in-the-loop (HITL) process \cite{amershi2014power} that derives candidate schemas from papers describing the same process and refines them with experts into reusable metadata models. This approach builds on information extraction (IE), Semantic Web technologies, LLM-based scientific knowledge representation \cite{martinezrodriguez2020information,shamsabadi-etal-2024-large,dagdelen2024structured,dsouza-etal-2025-mining}, and LLM-assisted schema discovery and generation \cite{mior2024large,zhang-etal-2025-schema,wu2025schema}. \SchemaMiner{} operationalizes the workflow by extracting, organizing, refining, and validating candidate properties under expert feedback \cite{schema-miner,schema-miner-pro}. \SciSchema{} scales not by replacing experts, but by shifting their starting point: \SchemaMiner{} surfaces candidate properties from many papers, while experts determine terminology, organization, coverage, and scientific adequacy. This workflow supported the expert curation of 16 schemas across five domains within two months, concentrating effort on scientific judgment rather than exhaustive manual extraction. This Data Descriptor releases and documents these schemas as reusable data resources.

%Introduce dataset descriptor
This Data Descriptor reports the first \SciSchema{} release: 16 community-curated scientific process schemas spanning five broad domains. The Biology and Biotechnology schemas cover Polymerase Chain Reaction (PCR), CRISPR-Cas9, Electrical Cell Stimulation, RNA-Seq Differential Gene Expression Analysis Workflow, and Bulk RNA-seq Library Preparation and Sequencing. The Materials and Chemistry schemas cover Fatigue Testing of Metallic Materials, Steam Reforming, Solvent Casting for Polymer Composites, Catalytic Pyrolysis of Mixed Plastic Waste, Metal-Organic Cage Synthesis, and Dynamic Mechanical Analysis (DMA). The Imaging and Measurement schemas cover Iterative Tomographic Image Reconstruction and Multi-Electrode Array (MEA) Recordings. The Physics schema covers Neutrino Event Reconstruction. The Psychology schemas cover the Stroop Task and the Torrance Tests of Creative Thinking (TTCT). For each process class, the released schema defines the structured attributes needed to describe process instances, including inputs, outputs, materials, instruments or software, parameters, settings, conditions, procedural steps, and provenance-relevant descriptors where applicable. This Data Descriptor describes the schema collection, its scope, its curation workflow, its representation in \SciSchema{} and ORKG templates, and its potential reuse in annotation, comparison, metadata enrichment, semantic publishing, and AI-ready scientific knowledge infrastructures. By presenting these resources as a documented dataset rather than as an experimental result, the paper provides a reusable reference point for researchers, curators, repository developers, and KG infrastructures that need structured descriptions of scientific processes.

\section*{Methods}

\subsection{Community formation and process selection}

The \SciSchema{} community was established through three calls for participation circulated on December 9, 2025, January 5, 2026, and January 24, 2026, respectively, via public mailing lists and professional networks. The calls targeted domain experts whose primary expertise lay outside computer science and invited individuals or small teams to propose experimental or simulation processes for structured, machine-assisted schema development. Participation required, at minimum, the provision of a sizeable corpus of full-text articles from which recurring process properties could be mined and expert feedback on the resulting schema structures; contributors could also participate directly in running the \SchemaMiner{} workflow. Interested contributors registered through an online form that collected their scientific field or discipline, a general process category, the proposed process name, a short lay description, and two representative scientific articles. The request for representative articles was intended to provide an initial check that the proposed process was clearly defined, documented in the scholarly literature, and suitable for schema mining.

\subsubsection{Process screening and selection}

The resulting submissions formed a candidate pool of 29 proposed processes spanning molecular biology, neuroscience, psychology, particle physics, astronomy, chemistry, materials science, engineering, artificial intelligence, computational imaging, and medicine. Each proposal was reviewed internally to determine its suitability for schema development.

Proposals were assessed according to three principal criteria. First, the proposed process was expected to represent an established scientific, experimental, observational, clinical, engineering, or analytical procedure supported by a sufficiently large and stable body of literature, typically extending over more than 15 years. This criterion was intended to ensure that the resulting schema would represent reusable community knowledge rather than a recently introduced or narrowly reported procedure. Second, the literature needed to contain a sufficiently large corpus of accessible publications describing executions of the process. Third, those publications needed to report recurring process-specific information at a level of detail that could support a substantive schema, including, where applicable, inputs, materials, instruments, software, parameters, operating conditions, procedural steps, measurements, and outputs.

The expected granularity of the resulting schema was therefore considered in addition to the size and maturity of the literature. This criterion was informed by previous applications of \SchemaMiner{} to Atomic Layer Deposition (ALD) \cite{ald1,ald2,ald3,schema-miner}, for which recurring process descriptions supported a schema containing more than 50 properties and nested structures. Preference was similarly given to processes characterized by numerous interrelated variables that recur across publications and are difficult to standardize through manual review alone.

Several proposed computational workflows were not selected despite being supported by substantial and mature publications. For example, proposals involving computational imaging and medical image classification commonly shared only a small number of broadly defined stages, such as data preparation, feature extraction, and classification. Although individual studies differed in the algorithms used, the recurring process structure was expected to yield a comparatively shallow schema dominated by high-level workflow steps. The selection decision therefore concerned the recurrence and granularity of process-level metadata rather than the scientific maturity or importance of the proposed field.

Application of these criteria resulted in the provisional acceptance of 18 processes, and the corresponding contributors were notified on January 31, 2026. Contributors for 16 processes subsequently confirmed their participation and proceeded to process specification, corpus collection, iterative schema mining, and expert validation.

\begin{figure}[!tb]
    \centering
    \includegraphics[width=\textwidth]{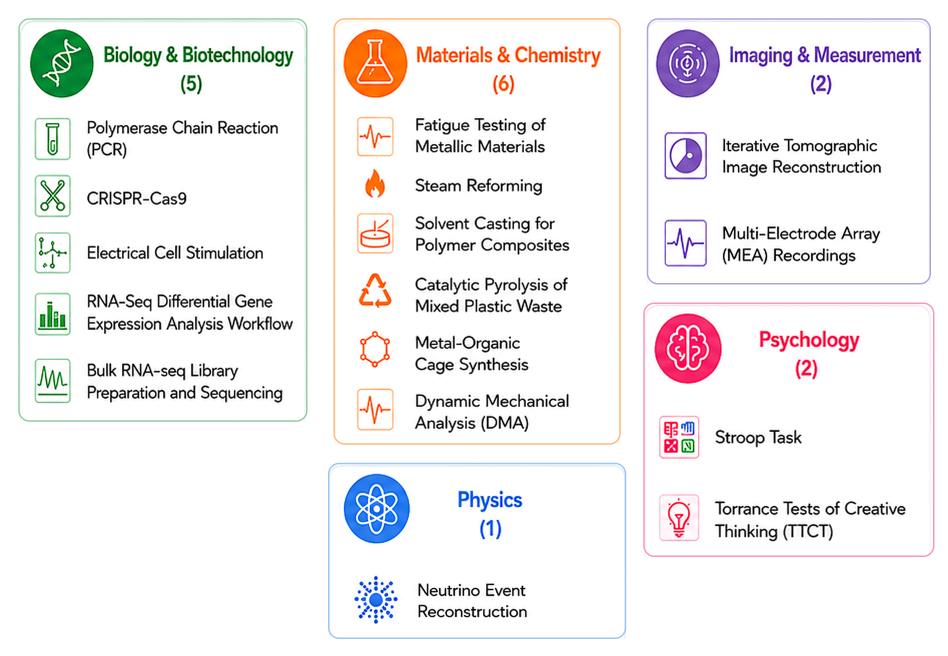}
    \caption{The first \SciSchema{} collection comprises 16 scientific process types organized into five broad disciplinary groups: Biology \& Biotechnology, Materials \& Chemistry, Imaging \& Measurement, Physics, and Psychology. Each process type is associated with a schema intended to capture recurring inputs, conditions, instruments, computational steps, measurements, and outputs reported in the scientific literature.}
    \label{fig:16-process-types}
\end{figure}

\subsection{Domain specification and source-paper collection}

The 16 confirmed processes were organized into five broad disciplinary groups: Biology \& Biotechnology, Materials \& Chemistry, Imaging \& Measurement, Physics, and Psychology, as shown in \autoref{fig:16-process-types}. This grouping was used for cataloging and presentation; schema development was conducted independently for each process. For each process, the domain expert prepared an informal specification document in PDF or Markdown format. The specification included a brief description of the process and an initial list of well known 5--15 properties to structure the process. These domain-informed specifications provided the input for initial schema generation.

The domain experts also assembled two collections of full-text scientific articles. The first was a small, manually curated set of approximately 10 papers selected for their direct relevance and detailed reporting of the process. The second was an expanded corpus of approximately 50 or more papers. This larger collection was screened for general relevance and document quality but was not subjected to the same level of manual curation. Its final size varied according to the availability and accessibility of suitable full-text publications. The process specification, curated paper set, and expanded corpus constituted the respective inputs to the three-stage \SchemaMiner{} workflow described below.

\subsection{Iterative schema-mining workflow}

The schemas were developed using the three-stage HITL workflow implemented in \SchemaMiner{} \cite{schema-miner,schema-miner-pro}. The source code is available under the MIT License at \url{https://github.com/sciknoworg/schema-miner}, and version 3.2.5 was used in this work. The library is distributed through PyPI at \url{https://pypi.org/project/schema-miner/}; the software can be installed using \texttt{pip install schema-miner}. Across the workflow, \SchemaMiner{} used LLMs to derive candidate JSON schema structures from domain specifications, paper evidence, and expert feedback. The papers provided repeated discourse-level evidence of how each process is reported in the literature, while domain experts guided the abstraction of these recurring properties into reusable schema structures with appropriate terminology, grouping, coverage, and scientific adequacy.

Twelve LLMs were applied independently to each process, yielding one three-stage schema-development trajectory per model. The set covered instruction following, reasoning orientation, and scale. Instruction-tuned models tested adherence to schema-design instructions and typed JSON output \cite{chung2024scaling}; reasoning-oriented models tested whether deliberative optimization produced different schema structures \cite{guo2025deepseek}; and small versus large models compared efficiency with expected in-context generalization \cite{hoffmann2022training}. Models with 8 billion parameters or fewer were classified as small.

The small instruction-tuned models were \modelref{https://huggingface.co/mistralai/Ministral-3-3B-Instruct-2512}{Ministral-3-3B-Instruct-2512}, \modelref{https://huggingface.co/mistralai/Ministral-3-8B-Instruct-2512}{Ministral-3-8B-Instruct-2512} \cite{liu2026ministral}, and \modelref{https://huggingface.co/Qwen/Qwen3-4B-Instruct-2507}{Qwen3-4B-Instruct-2507} \cite{qwen3technicalreport}; the large instruction-tuned models were \modelref{https://huggingface.co/google/gemma-3-27b-it}{Gemma-3-27B-IT} \cite{gemma_2025}, \modelref{https://huggingface.co/Qwen/Qwen3-30B-A3B-Instruct-2507}{Qwen3-30B-A3B-Instruct-2507}, \modelref{https://huggingface.co/Qwen/Qwen3-235B-A22B}{Qwen3-235B-A22B}, and \modelref{https://huggingface.co/mistralai/Mistral-Large-3-675B-Instruct-2512}{Mistral-Large-3-675B-Instruct-2512}; the small reasoning-oriented models were \modelref{https://huggingface.co/mistralai/Ministral-3-3B-Reasoning-2512}{Ministral-3-3B-Reasoning-2512}, \modelref{https://huggingface.co/mistralai/Ministral-3-8B-Reasoning-2512}{Ministral-3-8B-Reasoning-2512}, and \modelref{https://huggingface.co/Qwen/Qwen3-4B-Thinking-2507}{Qwen3-4B-Thinking-2507}; and the large reasoning-oriented models were \modelref{https://huggingface.co/deepseek-ai/DeepSeek-R1-Distill-Llama-70B}{DeepSeek-R1-Distill-Llama-70B} and \modelref{https://huggingface.co/Qwen/Qwen3-30B-A3B-Thinking-2507}{Qwen3-30B-A3B-Thinking-2507}.

Domain-expert feedback was collected according to written \href{https://github.com/sciknoworg/schema-miner/blob/main/assets/LLMs4SchemaDiscovery%20-%20Domain%20Expert%20Feedback%20Guidelines.pdf}{feedback guidelines}. For each reviewed schema, the feedback form recorded the scientific process and LLM model, followed by a free-text feedback field guided by questions on property merging, property grouping, missing properties, and adequacy of property descriptions. Experts could also provide direct schema edits, such as modified JSON fragments, and add further comments. The resulting feedback was supplied to the LLMs during the subsequent workflow stage or batch.

\textbf{Stage 1: initial schema generation.} Stage 1 produced the initial schema from the domain-expert specification alone, before introducing evidence from the literature. In \SchemaMiner{}, this step is implemented as a process-agnostic prompt template: the process name and short description are supplied as parameters, and the LLM is instructed to act as an expert schema designer and return a typed JSON schema. Other runtime settings, including the model and service provider, are configured through command-line arguments or the \texttt{.env} file. This design makes the workflow modular across all stages, allowing new processes, models, and providers to be substituted through configuration rather than code-level changes. The user prompt supplied the informal process specification and its initial list of 5--15 domain-informed properties. Each of the 12 models produced an independent candidate schema, which the domain experts reviewed to provide feedback for Stage 2.

\textbf{Stage 2: controlled paper set and expert feedback.} Stage 2 introduced the small, manually curated collection of approximately 10 papers for each process. The papers were divided into two batches and processed sequentially within each model trajectory. This batching enabled periodic, systematic expert intervention while keeping the review burden manageable. In principle, experts could review the updated schema after every paper; however, prior \SchemaMiner{} studies found that periodic feedback was similarly effective, given the strong instruction-following capabilities of contemporary LLMs \cite{schema-miner,schema-miner-pro}. At each iteration, the LLM received the next paper, the schema produced in the preceding iteration, and the most recent domain-expert feedback when available.

Feedback on the Stage 1 outputs was used during Stage 2 batch 1. Domain experts then reviewed the batch 1 schemas, and this feedback was used during batch 2. A second review was conducted after Stage 2 batch 2, providing the expert guidance used to initialize Stage 3.

\textbf{Stage 3: expanded corpus and iterative refinement.} Stage 3 applied the same iterative procedure to the expanded corpus of approximately 50 or more papers per process. The papers were again divided into two batches to allow periodic expert intervention during corpus-scale refinement. Stage 3 batch 1 began from the schemas produced at the end of Stage 2 and used the corresponding expert feedback. Domain experts then reviewed the batch 1 outputs, and this feedback was used during Stage 3 batch 2. The schemas produced after Stage 3 batch 2 constituted the final model-generated candidates for master-schema construction. Each process could therefore yield up to 12 final candidate schemas, one per model trajectory. Fewer outputs were available when a model did not return a valid schema, particularly when its context-processing capacity was insufficient for the accumulated schema and paper input.

\subsection{Master-schema construction and publication}

After Stage 3 batch 2, the model-generated schemas were treated as final candidate schemas rather than as final \SciSchema{} outputs. For each process, the domain expert received up to 12 candidate schemas, one from each model trajectory, and completed a final review form distinct from the iterative feedback forms used in the preceding stages. This final form asked the expert to rate each schema on a five-point Likert scale, from 1 (Worst) to 5 (Best), and to provide a master schema representing the preferred gold-standard structure for the process.

The master schema could be created by selecting one candidate schema as-is, editing a candidate schema, or combining elements from multiple model outputs. Experts were also asked to identify which model outputs informed the master schema and to describe how different parts of the final structure were derived. This step converted the model-generated candidates into expert-annotated schemas, with domain experts retaining control over terminology, grouping, coverage, nesting, and scientific adequacy.

The coordination team then reviewed the submitted master schemas for consistency of JSON structure, property organization, and formatting, while preserving the domain-expert decisions on process-specific content. The final schemas were published as the first \SciSchema{} collection through the \SciSchema{} website and associated repository. Where applicable, corresponding ORKG templates were created from the finalized schema structures so that the schema fields could be used for structured article-level descriptions of scientific processes. Detailed access information for the released files, repository, and template identifiers is provided in the Data Records and Data Availability sections.

\section*{Data Records}

The released data record is deposited in Zenodo as \emph{SciSchema.org: First Release Dataset} \cite{dsouza_2026_scischema}. It contains the first \SciSchema{} release as a versioned archival record, including the final expert-annotated schemas and the provenance materials needed to inspect how they were produced. The archive is organized around the 16 scientific processes rather than around file type, so that the final schema, intermediate schema-mining outputs, expert feedback, and source-paper metadata for a given process can be inspected together.

The high-level archive structure is shown below. The placeholder \texttt{<scientific-process>} denotes one of the 16 process-specific folders in the released \texttt{core/} directory.

\begin{quote}
\footnotesize
\begin{verbatim}
scischema.org/
|-- core/
|   |-- <scientific-process>/
|   |   |-- master-schema/
|   |   |   |-- master-schema.json
|   |   |   `-- master-schema.n3
|   |   |-- stage-1/
|   |   |-- stage-2/
|   |   |   |-- schema-batch1/
|   |   |   |-- schema-batch2/
|   |   |   |-- feedback-batch1/
|   |   |   `-- feedback-batch2/
|   |   |-- stage-3/
|   |   `-- metadata.csv
|-- raw-expert-feedback/
|-- community-building/
`-- paper-analysis/
\end{verbatim}
\end{quote}

\subsection*{Core process records}

The main data are provided in the \texttt{core/} directory, which contains one folder for each of the 16 scientific processes: Polymerase Chain Reaction (PCR), CRISPR-Cas9, Electrical Cell Stimulation, RNA-Seq Differential Gene Expression Analysis Workflow, Bulk RNA-seq Library Preparation and Sequencing, Fatigue Testing of Metallic Materials, Steam Reforming, Solvent Casting for Polymer Composites, Catalytic Pyrolysis of Mixed Plastic Waste, Metal-Organic Cage Synthesis, Dynamic Mechanical Analysis (DMA), Iterative Tomographic Image Reconstruction, Multi-Electrode Array (MEA) Recordings, Neutrino Event Reconstruction, Stroop Task, and Torrance Tests of Creative Thinking (TTCT). The folder names follow the labels used in the released archive and correspond to the processes shown in \autoref{fig:16-process-types}. 

Each process folder contains a \texttt{master-schema/} directory with the final expert-annotated schema in two machine-readable representations. The JSON Schema file \texttt{.json}, intended for document- and software-oriented reuse, defines the JSON structure for process instances, including metadata, nested property groups, terminal fields, data types, required fields, constraints, descriptions, examples, and reusable definitions where applicable. It supports record validation, data-entry form generation, constrained LLM or IE outputs, and software-pipeline integration. The SHACL file, serialized in Notation3 \texttt{.n3}, represents the same schema as node and property shapes with classes, predicates, labels, cardinality constraints, and ordering information. Intended for RDF and KG reuse, it supports graph validation, linked-data publication, and integration with infrastructures such as ORKG templates. Thus, both files encode the same schema-level content for different ecosystems: JSON Schema for structured records and SHACL/Notation3 for graph-based semantic representations.

\subsection*{Schema content and property organization}

The released schemas are schema-level models: they define the fields available for describing a scientific process but do not contain information extracted from individual articles. Applying a schema to a specific paper—that is, populating its fields from the reported content—is an IE task that the schemas are designed to support, but is not part of this release. Their property names and nesting define how future process instances can be described. Across the collection, the schemas organize process information into recurring functional blocks: identifying metadata, inputs or source materials, process settings and conditions, instruments or software, measurements or outputs, and analysis or quality-control information. The exact organization is process-specific and reflects the reporting structure of each domain.

For example, biology schemas distinguish between biological inputs, experimental design, wet-lab execution, and downstream analysis. In the CRISPR-Cas9 schema, fields such as \nolinkurl{inputs}, \nolinkurl{guideRNA}, \nolinkurl{sequence}, \nolinkurl{inSilicoDesigner}, \nolinkurl{modifications}, and \nolinkurl{nucleaseSource} identify the molecular components and design choices that determine the editing system. In RNA-seq schemas, blocks such as \nolinkurl{sample}, \nolinkurl{rna_extraction}, \nolinkurl{quality_control}, \nolinkurl{library_preparation}, \nolinkurl{sequencing}, and \nolinkurl{primary_data_processing} separate biological origin, RNA handling, library construction, sequencing configuration, and computational processing.

Materials and chemistry schemas use analogous nesting to capture the relation between material inputs, process conditions, and measured products. For catalytic pyrolysis, fields such as \nolinkurl{feedstock}, \nolinkurl{catalyst}, \nolinkurl{reactor}, \nolinkurl{process_conditions}, \nolinkurl{yields}, and \nolinkurl{analytical_characterisation} describe the plastic input stream, catalyst system, reactor configuration, reaction conditions, product distribution, and characterization methods. For fatigue testing, fields such as \nolinkurl{material}, \nolinkurl{specimenGeometry}, \nolinkurl{loadingParameters}, \nolinkurl{testEnvironment}, \nolinkurl{equipment}, and \nolinkurl{outputs} capture the tested alloy, specimen form, loading regime, environmental conditions, test setup, and fatigue-response measurements.

Schemas in imaging, physics, and psychology similarly encode domain-specific process structure. The CT-reconstruction schema separates acquisition settings, physical modeling, regularization, and solver configuration; the neutrino-reconstruction schema separates detector geometry, input data and detector models, reconstruction algorithms, and performance validation; and the Stroop-task schema separates participant and task metadata, stimuli, timing, experimental design, behavioral outputs, and analysis procedures. Thus, nesting is used not only for formatting but also to represent how process components depend on one another scientifically. The complete field definitions, constraints, descriptions, required fields, enumerations, and nested structures are provided in the corresponding \nolinkurl{master-schema/} files.

\subsection*{Schema-development provenance}

The process folders also preserve the schema-development trajectory. The \texttt{stage-1/}, \texttt{stage-2/}, and \texttt{stage-3/} directories contain the inputs and outputs from the successive \SchemaMiner{} stages, including process descriptions, model-generated schema files, and expert-feedback files per model. For batched stages, as applicable, outputs are organized into batch-specific subdirectories such as \texttt{schema-batch1/}, \texttt{schema-batch2/}, \texttt{feedback-batch1/}, and \texttt{feedback-batch2/}. These records make the development history of each master schema inspectable: users can compare the final expert-annotated schema with earlier model-generated candidates and the expert feedback that guided subsequent iterations.

Each process folder also contains a \texttt{metadata.csv} file describing the source-paper corpus used during schema mining. The file records the source publication title (\texttt{paper\_title}), DOI or URL (\texttt{doi\_or\_url}), and assigned workflow stage (\texttt{assigned\_stage}). Source-paper PDFs are not redistributed because licensing and redistribution conditions vary across publishers and venues. The schema-mining runs relied on lawful access to the articles, including institutional access and open-access sources where available. The public record therefore provides bibliographic source metadata and derived schema-mining outputs, rather than original article files or substantial reproduced article text. This allows users to reconstruct the corpus where access rights permit and rerun \SchemaMiner{} stage by stage using the same paper-to-stage assignments.

\subsection*{Community, feedback, and analysis records}

The archive also includes cross-process materials documenting the broader community-based schema-development workflow. The \texttt{raw-expert-feedback/} directory contains exported expert-feedback form responses in TSV format, preserving the anonymized human-review data collected during iterative schema refinement and final master-schema construction. The \texttt{community-building/} directory contains the calls for participation, contributor instructions, and schema-development or feedback guidelines used during community data collection. The \texttt{paper-analysis/} directory contains the data, scripts, and generated figures used to reproduce the analyses reported in this Data Descriptor.

Together, these records support three levels of reuse. Users interested only in applying the released schemas can use the \texttt{master-schema/} files. Users interested in provenance can inspect the intermediate stage outputs and expert-feedback records. Users interested in reproducing or extending the analyses in this paper can use the \texttt{paper-analysis/} directory together with the process-level \texttt{metadata.csv} files and released master schemas.

\begin{comment}

this section needs to contain discussions on the topics mentioned here

all relevant lines of work, some of which are already mentioned above

schema.org

JSON-LD

(for semantics)

RO-Crate

(package schemas with metadata and provenance)

PROV-O

(represent provenance chains)

SHACL

(validation)

LinkML

(generation and validation)

Bioschemas

(domain examples)

ORKG templates

(already close to your vision)

\end{comment}

%What can we do with these schemas: 1) data can become discoverable on the web via major search engines like google, bing etc., 2) can be exploited within our informatics communities like DPC, FAIR, NFDI4DS ... identify all NFDIs (this needs to be fleshed out in terms of what it means say for FAIRMat or NFDI4BioDiv... map a roadmap diagram with all the NFDI consortiums considered), (these are just examples for how bioschemas promotes itself and envisions its integration: can be fed into data aggregators such as identifiers.org, FAIRDARE, OrphaNet or the Covid-19 data portal), and 3) use to feed into large knowledge graphs such as datacite PID graph, open research knowledge graph, openaire research graph, researchgraph.org .. then this becomes part of a larger collection of knowledge about the scientific processes and can be exploited more and more widely.

% Optional: remove this section if not needed.
\section*{Data Overview}

\begin{figure}[!tb]
\centering
\includegraphics[width=0.8\textwidth]{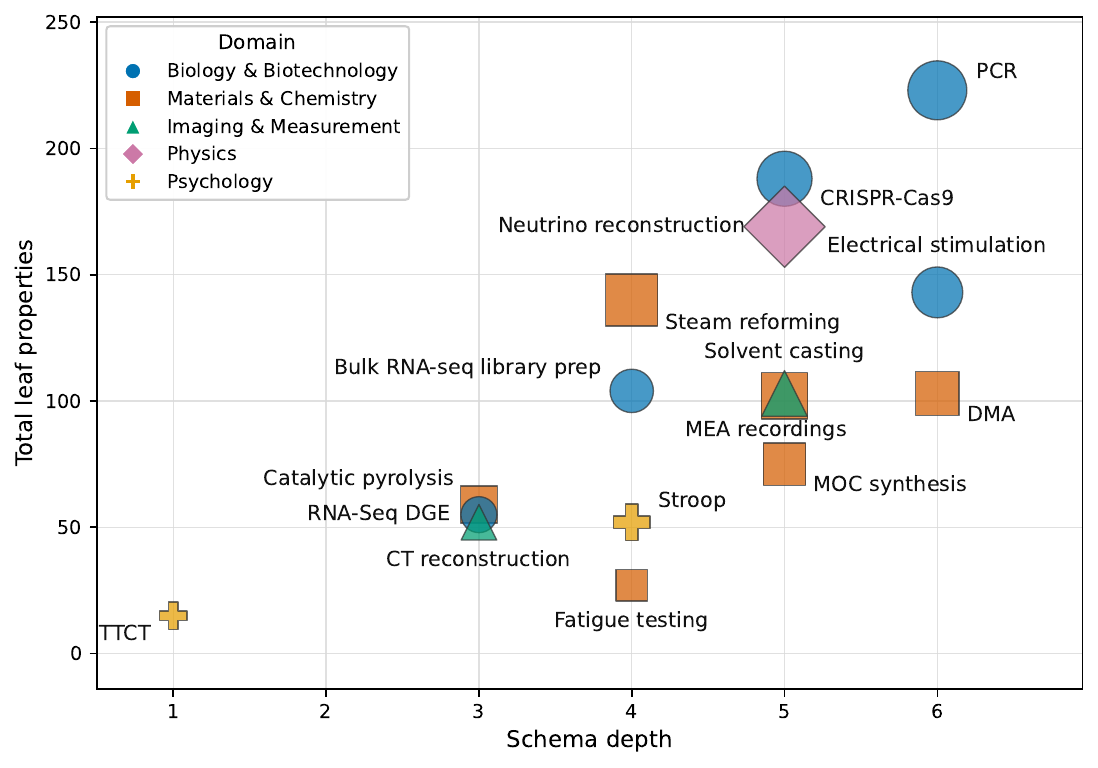}
\caption{\textbf{Structural overview of the released scientific-process schemas.} Each point represents one final expert-annotated schema. The x-axis shows schema depth, the y-axis shows the number of leaf properties, marker size reflects the total number of properties, and marker color and shape indicate the broad scientific domain. Abbreviations: PCR, polymerase chain reaction; DMA, dynamic mechanical analysis; MOC, metal-organic cage; MEA, multi-electrode array; RNA-Seq DGE, RNA sequencing differential gene expression; CT, computed tomography; TTCT, Torrance Tests of Creative Thinking.}
\label{fig:schema-structural-overview}
\end{figure}

The released collection comprises 16 expert-annotated scientific-process schemas spanning five broad domains: Biology \& Biotechnology, Materials \& Chemistry, Imaging \& Measurement, Physics, and Psychology. \autoref{fig:schema-structural-overview} gives a compact view of the structural range represented in the dataset. Across the collection, schema depth ranges from 1 to 6 levels, total properties from 15 to 285, and leaf properties from 15 to 223. This variation indicates that the collection includes both compact schemas for more bounded process descriptions and larger schemas with many terminal fields for processes that require more fine-grained reporting. The figure is intended as an overview of schema structure only; the corresponding tabular structural summary is provided with the released data files for users who wish to inspect or reuse the individual schema statistics.

\section*{Technical Validation}
%Explain how the reliability, consistency, completeness, and/or accuracy of the dataset were assessed. Describe validation experiments, benchmark comparisons, expert review, error analysis, schema validation, reproducibility checks, or sanity checks.

The technical validation comprises four complementary analyses. \autoref{fig:paper-length} and \autoref{fig:feedback-length} characterize input heterogeneity and expert-feedback behavior across processes, models, and stages. \autoref{fig:schema-length} uses schema-token length as a coarse proxy for output scale, while \autoref{fig:schema-progression} and \autoref{fig:schema-summary} assess structural schema complexity using properties, leaf nodes, and schema levels. Finally, \autoref{fig:final-validation} evaluates the final Stage 3 batch 2 schema candidates through domain-expert ratings and master-schema source selection, linking model-generated outputs to the expert-annotated \SciSchema{} schemas.

\subsection{Input heterogeneity and expert-review coverage}

\begin{figure}[!tb]
    \centering
    \includegraphics[width=\textwidth]{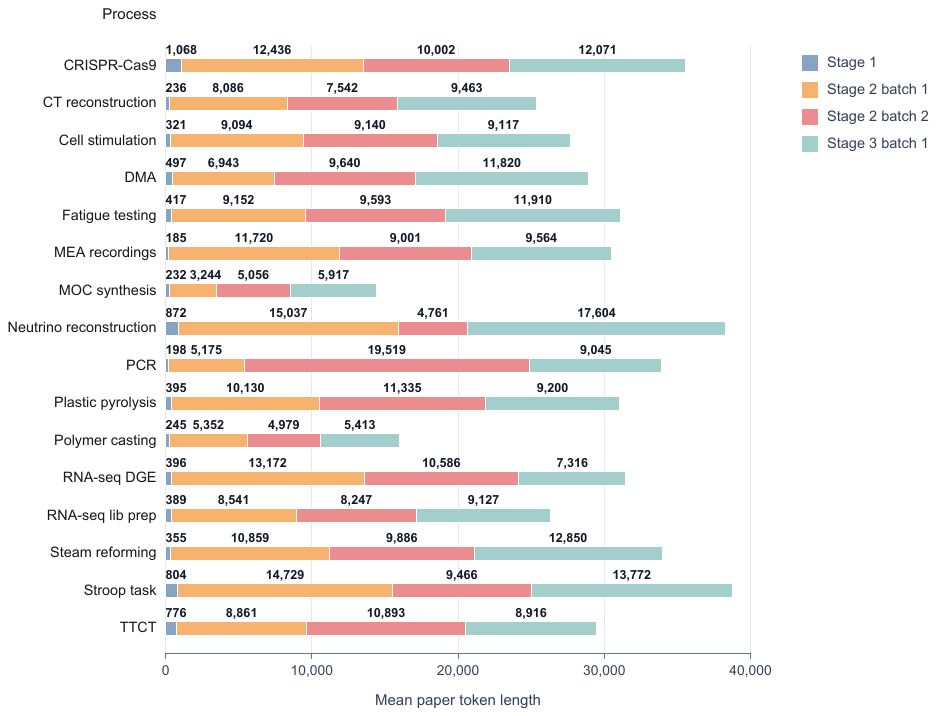}
    \caption{Average paper token length by process, aggregated across stage and batch groups. Error bars indicate the minimum and maximum paper token lengths.}
    \label{fig:paper-length}
\end{figure}

\begin{figure}[!tb]
    \centering
    \includegraphics[width=\textwidth]{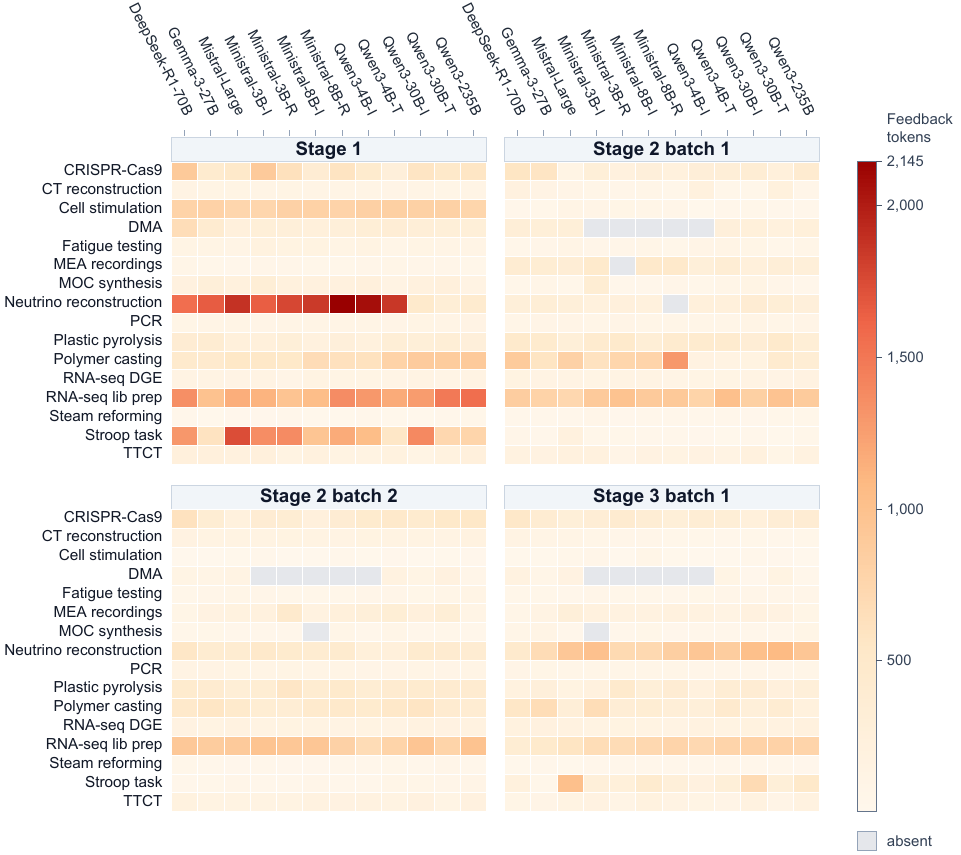}
    \caption{\textbf{Domain-expert feedback length across processes, models, and \SchemaMiner{} stages.} Heatmaps show the average length of domain-expert feedback, measured in whitespace-separated tokens, for each process–model combination. Feedback was provided on the schema output of a given stage and used to guide schema mining in the subsequent stage. Rows show shortened process names and columns show shortened model names. Separate panels show feedback lengths for Stage 1, Stage 2 batch 1, Stage 2 batch 2, and Stage 3 batch 1. Darker cells indicate longer expert feedback and lighter cells the opposite. Grey cells indicate absent feedback data. Model abbreviations are as follows: DeepSeek-R1-70B = deepseek-r1-distill-llama-70b; Gemma-3-27B = gemma-3-27b-it; Mistral-Large = mistral-large-3-675b-instruct; Ministral-3B-I = Ministral-3-3B-Instruct; Ministral-3B-R = Ministral-3-3B-Reasoning; Ministral-8B-I = Ministral-3-8B-Instruct; Ministral-8B-R = Ministral-3-8B-Reasoning; Qwen3-4B-I = Qwen3-4B-Instruct; Qwen3-4B-T = Qwen3-4B-Thinking; Qwen3-30B-I = qwen3-30b-a3b-instruct; Qwen3-30B-T = qwen3-30b-a3b-thinking; and Qwen3-235B = qwen3-235b-a22b.}
    \label{fig:feedback-length}
\end{figure}

This part of the technical validation examines the heterogeneity of the inputs used across the schema-mining workflow and the extent of domain-expert guidance provided across processes, models, and stages. As shown in \autoref{fig:paper-length}, the Stage 1 process specifications were consistently shorter than the scientific articles introduced during Stages 2 and 3, while article lengths also varied substantially across processes and batches. The workflow was therefore applied to source materials with differing input scales rather than to uniformly sized documents.

Domain experts reviewed the generated schemas after the designated stages and batches, and their written feedback was used to guide the subsequent refinement step. \autoref{fig:feedback-length} shows the coverage and length of these feedback records across processes, models, and workflow stages. Feedback length indicates the extent of human guidance provided after inspection of each schema output and therefore where more extensive expert intervention was required before the workflow proceeded. Variation was more pronounced across scientific processes than across models. Feedback was longest after Stage 1, averaging 460.5 tokens, whereas the subsequent feedback rounds were similar in length, averaging 243.9 tokens after Stage 2 batch 1, 239.4 after Stage 2 batch 2, and 245.0 after Stage 3 batch 1. Stage 3 batch 2 is not included because it was the final generation stage and was not followed by another feedback round.

\subsection{Schema-token length as a coarse output-scale proxy}

\begin{figure}[!tb]
    \centering
    \includegraphics[width=\textwidth]{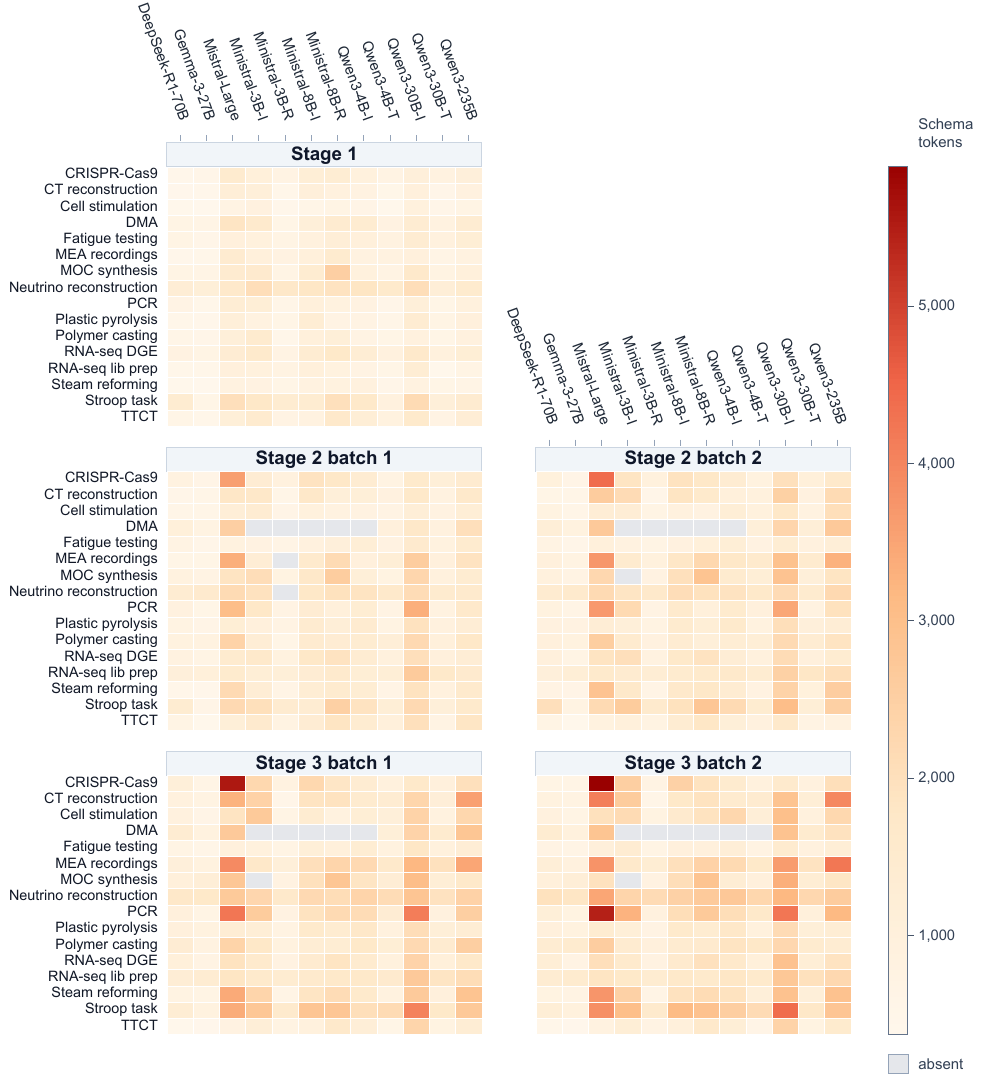}
    \caption{\textbf{Schema length across processes, models, and workflow stages.}
    Heatmaps show the average length of generated JSON schemas, measured in whitespace-separated tokens, for each process--model combination. Rows represent shortened process names and columns represent shortened model names. Separate panels show Stage 1, Stage 2 batches 1 and 2, and Stage 3 batches 1 and 2. Darker cells indicate longer schemas and lighter cells indicate shorter schemas; grey cells indicate that the model did not produce a valid schema for the corresponding process. Model abbreviations are as follows: DeepSeek-R1-70B = deepseek-r1-distill-llama-70b; Gemma-3-27B = gemma-3-27b-it; Mistral-Large = mistral-large-3-675b-instruct; Ministral-3B-I = Ministral-3-3B-Instruct; Ministral-3B-R = Ministral-3-3B-Reasoning; Ministral-8B-I = Ministral-3-8B-Instruct; Ministral-8B-R = Ministral-3-8B-Reasoning; Qwen3-4B-I = Qwen3-4B-Instruct; Qwen3-4B-T = Qwen3-4B-Thinking; Qwen3-30B-I = qwen3-30b-a3b-instruct; Qwen3-30B-T = qwen3-30b-a3b-thinking; and Qwen3-235B = qwen3-235b-a22b.}
    \label{fig:schema-length}
\end{figure}

\autoref{fig:schema-length} provides an output-level sanity check based on the length of the generated JSON schemas. Schema length was measured as the number of whitespace-separated tokens and used as a coarse indicator of output scale, because schema expansion may introduce additional properties, nested structures, descriptions, and terminal IE targets. Averaged across all processes and models, schema length increased from 983.8 tokens in Stage 1 to 1321.5 in Stage 2 batch 1, 1484.7 in Stage 2 batch 2, 1679.4 in Stage 3 batch 1, and 1791.0 in Stage 3 batch 2.

All 12 models produced longer schemas on average in Stage 3 batch 2 than in Stage 1. At the process level, 13 of the 16 processes showed a monotonic increase across successive stages and batches; Fatigue testing, MOC synthesis, and TTCT contained one or more local decreases. These trajectories show that schema development generally increased output scale as source-paper evidence and expert guidance were incorporated, while also permitting consolidation or restructuring between successive stages. Because token length alone cannot distinguish structural expansion from verbose descriptions, property naming, or repeated formatting, the following analysis examines corresponding changes in direct structural measures, including properties, leaf nodes, and schema levels.

\subsection{Structural schema complexity beyond token length}

\begin{figure}[!tb]
\centering
\includegraphics[width=\textwidth]{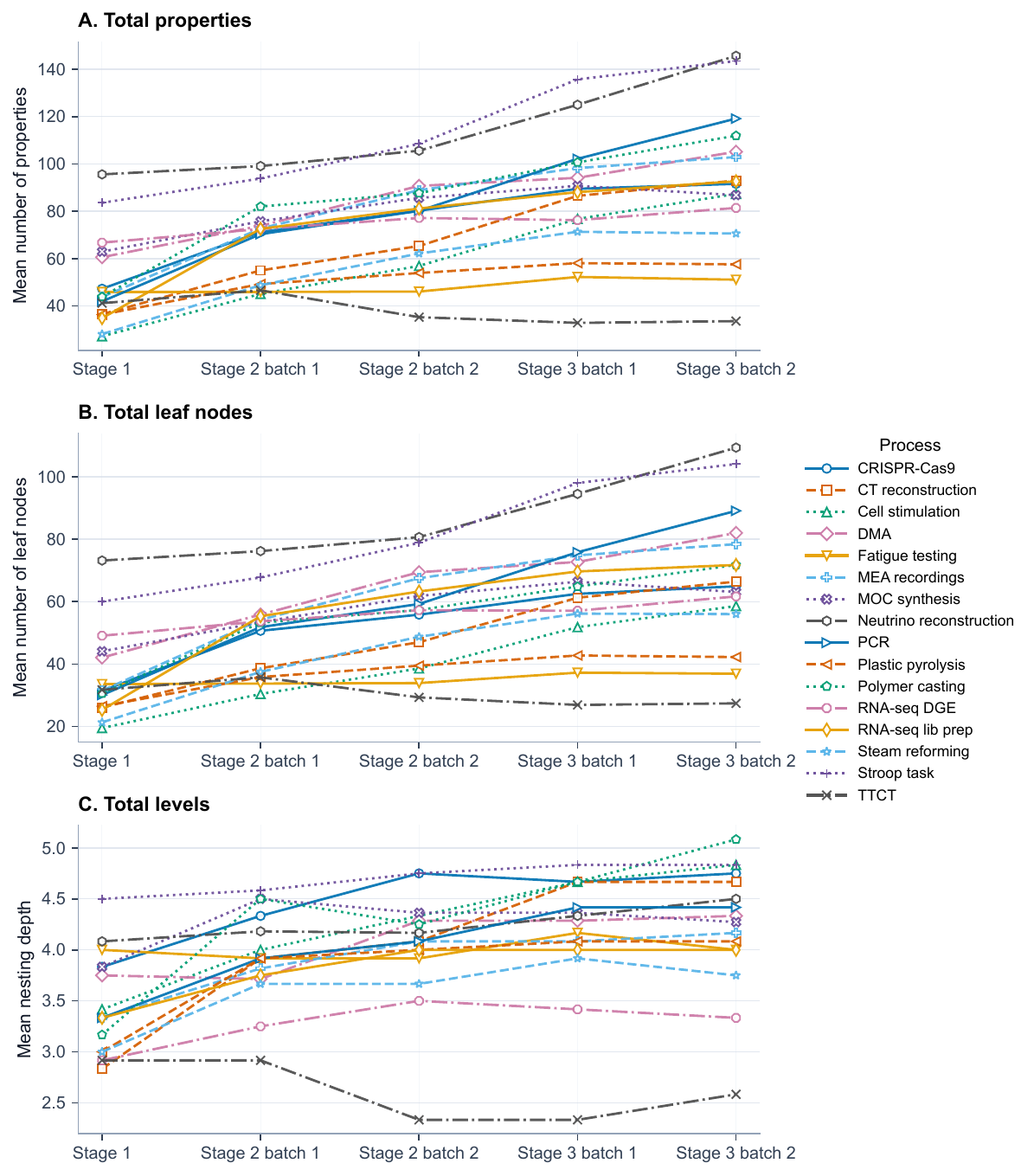}
\caption{\textbf{Process-level progression of schema structural complexity across stages and batches.} Lines show process-wise means across models for total properties, total leaf nodes, and total levels. Later stages and batches generally yield more structurally detailed schemas, although trajectories vary across processes.}
\label{fig:schema-progression}
\end{figure}

\begin{figure}[!tb]
\centering
\includegraphics[width=\textwidth]{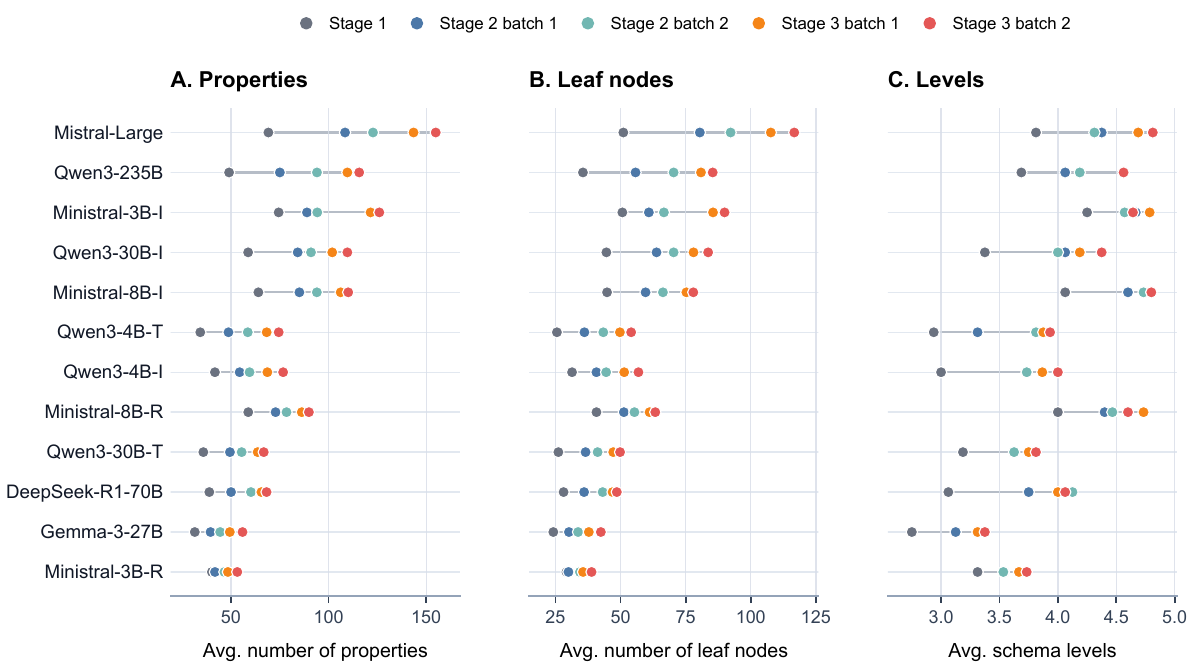}
\caption{\textbf{Model-level progression of schema structural complexity across stages and batches.} Panels show model-wise means for total properties, leaf nodes, and schema levels, aggregated across processes. Each point represents a model at one stage/batch setting. Models are ordered by overall structural expansion, with larger rightward shifts indicating greater increases. Model abbreviations are as follows: DeepSeek-R1-70B = deepseek-r1-distill-llama-70b; Gemma-3-27B = gemma-3-27b-it; Mistral-Large = mistral-large-3-675b-instruct; Ministral-3B-I = Ministral-3-3B-Instruct; Ministral-3B-R = Ministral-3-3B-Reasoning; Ministral-8B-I = Ministral-3-8B-Instruct; Ministral-8B-R = Ministral-3-8B-Reasoning; Qwen3-4B-I = Qwen3-4B-Instruct; Qwen3-4B-T = Qwen3-4B-Thinking; Qwen3-30B-I = qwen3-30b-a3b-instruct; Qwen3-30B-T = qwen3-30b-a3b-thinking; and Qwen3-235B = qwen3-235b-a22b.}
\label{fig:schema-summary}
\end{figure}

To test whether schema-token length reflected structural growth rather than only verbosity, we evaluated three metrics: total properties, total leaf nodes, and total levels. Properties measure schema size, leaf nodes approximate terminal IE targets, and levels measure nesting depth. Schema-token length correlated strongly with properties ($r=0.898$) and leaf nodes ($r=0.915$), and moderately with total levels ($r=0.556$), supporting its use as a coarse proxy while confirming the need for direct structural analysis.

As shown in \autoref{fig:schema-progression}, the structural metrics show clear stage-wise expansion, although trajectories varied by process. Averaged across all processes and models, properties increased from 49.8 in Stage 1 to 66.8, 74.9, 85.9, and 91.7 across Stage 2 batch 1 through Stage 3 batch 2. Leaf nodes similarly increased from 36.0 to 48.6, 55.1, 63.0, and 67.3, while total levels rose from 3.45 to 3.94, 4.03, 4.18, and 4.22. All 12 models had higher mean values for all three metrics in Stage 3 batch 2 than in Stage 1; between these stages, 15 of the 16 processes increased in properties and leaf nodes, and 14 increased in total levels. Strict monotonic growth across every adjacent stage/batch occurred for 10 processes in properties and leaf nodes and for 7 in total levels. TTCT became more compact, decreasing from 41.2 properties and 31.8 leaf nodes to 33.6 and 27.4, respectively, while Fatigue testing was comparatively flat, increasing from 45.9 to 51.1 properties and from 33.6 to 36.9 leaf nodes, with no net increase in total levels. By contrast, PCR expanded from 41.9 to 119.2 properties and from 30.1 to 89.1 leaf nodes. Solvent casting, cell stimulation, MEA recordings, RNA-seq library preparation, and CT reconstruction also showed large increases, indicating that later paper batches produced the strongest expansion when they introduced additional procedural distinctions and extraction targets.

As shown in \autoref{fig:schema-summary}, model-level differences were also pronounced. \href{https://huggingface.co/mistralai/Mistral-Large-3-675B-Instruct-2512}{mistral-large-3-675b-instruct} produced the largest schemas on average, with 119.8 properties and 89.6 leaf nodes, followed by \href{https://huggingface.co/mistralai/Ministral-3-3B-Instruct-2512}{Ministral-3-3B-Instruct} with 100.1 properties and 70.0 leaf nodes. The most compact schemas came from \href{https://huggingface.co/google/gemma-3-27b-it}{gemma-3-27b-it} and \href{https://huggingface.co/mistralai/Ministral-3-3B-Reasoning-2512}{Ministral-3-3B-Reasoning}, with 44.2 and 46.1 properties, respectively, and 33.7 leaf nodes each. Instruction-oriented models were more expansive than thinking/reasoning-oriented models, averaging 84.8 versus 58.1 properties, 62.3 versus 42.0 leaf nodes, and 4.1 versus 3.8 levels. Model scale was less explanatory: large models averaged 75.4 properties, 56.3 leaf nodes, and 3.9 levels, compared with 71.9 properties, 51.5 leaf nodes, and 4.1 levels for small models. Neutrino reconstruction and Stroop task produced the largest schemas by process, averaging 114.4 and 113.0 properties and 87.0 and 81.7 leaf nodes, respectively. TTCT, Fatigue testing, Plastic pyrolysis, and Steam reforming produced the smallest schemas by property count.

Schema depth followed a related but distinct pattern. The deepest schemas came from \href{https://huggingface.co/mistralai/Ministral-3-8B-Instruct-2512}{Ministral-3-8B-Instruct} and \href{https://huggingface.co/mistralai/Ministral-3-3B-Instruct-2512}{Ministral-3-3B-Instruct}, with mean depths of 4.59 and 4.58 levels, followed by \href{https://huggingface.co/mistralai/Ministral-3-8B-Reasoning-2512}{Ministral-3-8B-Reasoning} (4.43) and \href{https://huggingface.co/mistralai/Mistral-Large-3-675B-Instruct-2512}{mistral-large-3-675b-instruct} (4.40). The shallowest came from \href{https://huggingface.co/google/gemma-3-27b-it}{gemma-3-27b-it} (3.18), \href{https://huggingface.co/mistralai/Ministral-3-3B-Reasoning-2512}{Ministral-3-3B-Reasoning} (3.55), \href{https://huggingface.co/Qwen/Qwen3-4B-Thinking-2507}{Qwen3-4B-Thinking} (3.57), and \href{https://huggingface.co/Qwen/Qwen3-30B-A3B-Thinking-2507}{qwen3-30b-a3b-thinking} (3.60). At the process level, Stroop task had the deepest schemas on average (4.70 levels), followed by CRISPR-Cas9 (4.47), Polymer casting (4.33), MOC synthesis (4.26), and Neutrino reconstruction (4.25), whereas TTCT averaged 2.62 levels. Lower depth does not itself indicate lower quality, since some processes require fewer nested distinctions or terminal fields.

Leaf-node counts further distinguish breadth from depth. The highest individual counts occurred in late-stage \href{https://huggingface.co/mistralai/Mistral-Large-3-675B-Instruct-2512}{mistral-large-3-675b-instruct} schemas: CRISPR-Cas9 in Stage 3 batch 2 had 235 leaf nodes and PCR had 230. Overall, \href{https://huggingface.co/mistralai/Mistral-Large-3-675B-Instruct-2512}{mistral-large-3-675b-instruct} produced the largest average numbers of properties and leaf nodes and ranked first in leaf-node count for 9 of the 16 process-level averages. No single model dominated schema depth, which was more distributed across the Ministral and Mistral variants. Thus, breadth and depth capture complementary dimensions of schema complexity, and compact schemas may reflect process-specific adaptation rather than insufficient generation.

Finally, structural complexity was not simply explained by input paper length. At the row level, Pearson correlations between mean paper length and total properties, leaf nodes, and total levels were modest: $r=0.255$, $r=0.277$, and $r=0.198$, respectively. After averaging over process-stage-batch groups, they increased to $r=0.401$, $r=0.435$, and $r=0.332$, partly because Stage 1 inputs and schemas were both shorter. Excluding Stage 1, the row-level correlations became weak: $r=0.033$ for properties, $r=0.066$ for leaf nodes, and $r=-0.021$ for levels. Thus, schema-token length was strongly related to structural complexity, but input paper length alone was not; structural variation instead reflected process-specific content, evidence accumulation, expert-guided refinement, and model-specific generation behavior.

\subsection{Master expert-annotated schemas}

\begin{figure}[!tb]
\centering
\includegraphics[width=0.8\textwidth]{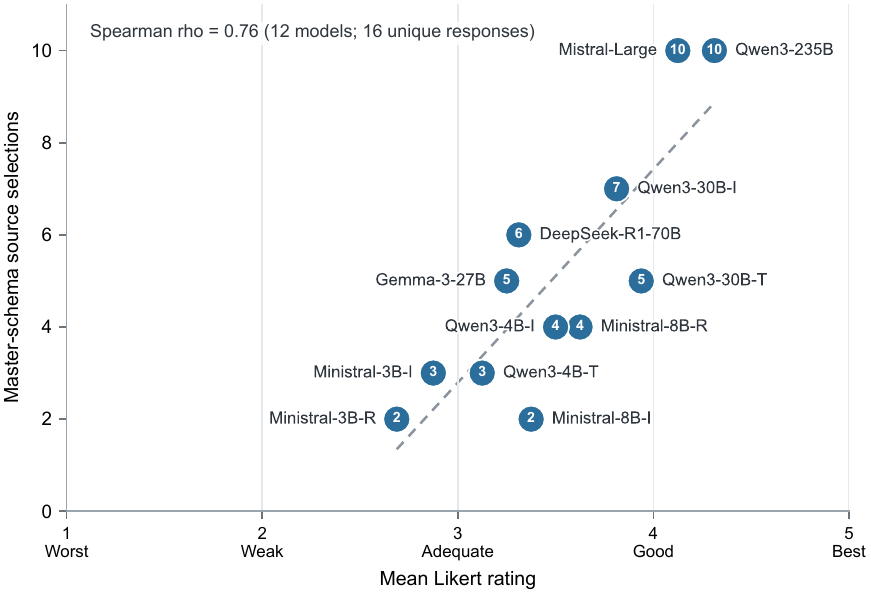}
\caption{\textbf{Relationship between final expert ratings and master-schema source selection.} Each point represents one Stage 3 batch 2 LLM, with mean expert Likert rating on the x-axis and the number of master-schema source selections on the y-axis. Point labels show source-selection counts. The dashed line indicates the positive rank association between rating and reuse in master-schema construction (Spearman $\rho=0.76$ across 12 models and 16 process responses). Model abbreviations are as follows: DeepSeek-R1-70B = deepseek-r1-distill-llama-70b; Gemma-3-27B = gemma-3-27b-it; Mistral-Large = mistral-large-3-675b-instruct; Ministral-3B-I = Ministral-3-3B-Instruct; Ministral-3B-R = Ministral-3-3B-Reasoning; Ministral-8B-I = Ministral-3-8B-Instruct; Ministral-8B-R = Ministral-3-8B-Reasoning; Qwen3-4B-I = Qwen3-4B-Instruct; Qwen3-4B-T = Qwen3-4B-Thinking; Qwen3-30B-I = qwen3-30b-a3b-instruct; Qwen3-30B-T = qwen3-30b-a3b-thinking; and Qwen3-235B = qwen3-235b-a22b.}
\label{fig:final-validation}
\end{figure}

The final validation step assessed whether domain experts' ratings of the Stage 3 batch 2 outputs were reflected in the schemas they selected, edited, or combined when constructing the final master schemas. The final feedback set comprised 16 process responses, each rating 12 model-generated schemas on a five-point Likert scale: 1 (Worst), 2 (Weak), 3 (Adequate), 4 (Good), and 5 (Best). Overall, 55.2\% of ratings were 4 or 5, and 20.3\% were 5. The most highly rated candidates were \href{https://huggingface.co/Qwen/Qwen3-235B-A22B}{qwen3-235b-a22b} (mean 4.31; 81.3\% Good/Best), \href{https://huggingface.co/mistralai/Mistral-Large-3-675B-Instruct-2512}{mistral-large-3-675b-instruct-2512} (4.13; 75.0\%), \href{https://huggingface.co/Qwen/Qwen3-30B-A3B-Thinking-2507}{qwen3-30b-a3b-thinking-2507} (3.94; 68.8\%), and \href{https://huggingface.co/Qwen/Qwen3-30B-A3B-Instruct-2507}{qwen3-30b-a3b-instruct-2507} (3.81; 68.8\%).

As shown in \autoref{fig:final-validation}, higher-rated outputs were more frequently reused during master-schema construction, with a positive rank association between mean expert rating and source-selection frequency (Spearman $\rho=0.76$ across 12 models). \href{https://huggingface.co/Qwen/Qwen3-235B-A22B}{qwen3-235b-a22b} and \href{https://huggingface.co/mistralai/Mistral-Large-3-675B-Instruct-2512}{mistral-large-3-675b-instruct-2512} were each selected 10 times, followed by \href{https://huggingface.co/Qwen/Qwen3-30B-A3B-Instruct-2507}{qwen3-30b-a3b-instruct-2507} with 7 selections and \href{https://huggingface.co/deepseek-ai/DeepSeek-R1-Distill-Llama-70B}{deepseek-r1-distill-llama-70b} with 6. Among responses containing explicit source-model information, selected outputs had higher mean ratings than non-selected outputs (4.31 versus 3.11), as well as higher Good/Best rates (85.5\% versus 40.8\%) and Best-rating rates (50.0\% versus 6.2\%).

These results show that master-schema construction drew primarily on expert-preferred model outputs while retaining process-specific domain judgment. The final master schemas are therefore expert-annotated consolidations of selected Stage 3 candidates rather than unmodified model outputs.

\subsection{JSON Schema and SHACL shape-graph validation}

The final schema release was checked in both JSON Schema and SHACL formats. The 16 JSON Schema files were parsed as JSON and validated against the JSON Schema Draft 2020-12 ( \url{https://json-schema.org/draft/2020-12}) meta-schema, the draft used by the collection. Each schema was checked for a Draft 2020-12 declaration, syntactic validity, root-level title and description metadata, version value \texttt{1.0.0}, and stable \texttt{\$id} values declared under the \SciSchema{} schema namespace. Filename stems were also compared with the corresponding schema slugs encoded in the \texttt{\$id} values.

The corresponding 16 SHACL files were validated as RDF shape graphs serialized in Notation3 and checked against the W3C SHACL Recommendation (\url{https://www.w3.org/TR/shacl/}). Each \texttt{.n3} file in the process-level \texttt{master-schema/} directory was parsed with \href{https://rdflib.readthedocs.io/en/stable/}{RDFLib} \cite{Krech_RDFLib_2026}, a Python library for processing RDF graphs, checked for SHACL node-shape and property-shape declarations, and evaluated with \href{https://github.com/RDFLib/pySHACL}{pySHACL} \cite{Sommer_pySHACL_2026}, a Python SHACL validator, to confirm that the shape graph was well formed. The validation also checked datatype lexical consistency for numeric bounds used in SHACL constraints.

All 16 JSON Schema files passed Draft 2020-12 meta-schema validation. All 16 SHACL \texttt{.n3} files parsed successfully, contained SHACL shape declarations, and passed SHACL meta-validation. Targeted regression checks confirmed that previously identified issues had been resolved, including non-negative Kelvin lower bounds, corrected MEA quantity-object constraints, removal of stale Draft 07 wording, corrected electrical-cell-stimulation descriptions, inclusion of the RNA-seq differential-expression root description, and consistent process-level schema titles.

% Optional: remove this section if not needed.
\section*{Usage Notes}

The released schemas provide standards-based structural representations for scientific-process metadata. The JSON Schema representations can be used to validate process records, generate data-entry forms, and constrain LLM or other IE outputs by specifying the fields, datatypes, required elements, cardinalities, and permitted values or ranges. In particular, they can serve as target structures in schema-guided IE pipelines operating over scientific publications, a widely used approach in which relevant information is extracted from article text, tables, figures, or supplementary materials into predefined fields and subsequently checked for structural conformity. The SHACL representations support validation and linked-data publication of RDF process records and their integration into KG infrastructures, including through the released ORKG templates. By making protocols, materials, instruments, software, conditions, measurements, and outputs explicit, instantiated records can support metadata standardization, quality-control checks, cross-study comparison, and machine-assisted workflows. Broader integration with domain research-data infrastructures, PID and research-graph services, and Web discovery systems would require mappings to relevant vocabularies and infrastructure-specific profiles. Such mappings constitute a future community-development step built upon the present schema release.

\section*{Data Availability}

The resource described in this Data Descriptor is deposited in Zenodo as \emph{SciSchema.org: First Release Dataset} under DOI \url{https://doi.org/10.5281/zenodo.21254268} \cite{dsouza_2026_scischema} and is released under CC BY-SA 4.0. The Zenodo record provides the versioned archival release of the final expert-annotated schemas, intermediate schema-development records, expert-feedback files, source-paper metadata, community materials, and analysis files described in the Data Records section. No authentication is required to access the public dataset files. The original source-paper PDFs are not redistributed; access to the source publications depends on the user's institutional subscriptions, open-access availability, or other applicable access rights.

A live version of the resource is maintained at \url{https://scischema.org/}. Persistent access is provided through the W3ID namespace at \url{https://w3id.org/scischema/}, with the schema catalog available at \url{https://w3id.org/scischema/schemas}. The development repository containing the JSON schema files served by the website is available at \url{https://github.com/scischema/website/tree/main/schemas}. Individual schema landing pages and corresponding ORKG template identifiers are listed in \autoref{tab:data-availability-schema-links}.

\begin{table}[!tb]
\centering
\footnotesize
\caption{Individual \SciSchema{} schema landing pages and corresponding ORKG template identifiers. Schema paths resolve under \url{https://w3id.org/scischema/schema/}; ORKG template identifiers resolve under \url{https://orkg.org/templates/}.}
\label{tab:data-availability-schema-links}
\begin{tabular}{p{0.4\textwidth} p{0.315\textwidth} p{0.2\textwidth}}
\hline
\textbf{Scientific process} & \textbf{\SciSchema{} schema path} & \textbf{ORKG template ID} \\
\hline
Polymerase Chain Reaction (PCR) & \href{https://w3id.org/scischema/schema/biology/pcr}{\url{biology/pcr}} & \href{https://orkg.org/templates/R1640868}{R1640868} \\
CRISPR-Cas9 & \href{https://w3id.org/scischema/schema/biology/crispr-cas9/}{\url{biology/crispr-cas9}} & \href{https://orkg.org/templates/R1900916}{R1900916} \\
Electrical Cell Stimulation & \href{https://w3id.org/scischema/schema/biology/electrical-cell-stimulation/}{\url{biology/electrical-cell-stimulation}} & \href{https://orkg.org/templates/R1899683}{R1899683} \\
RNA-Seq Differential Gene Expression Analysis Workflow & \href{https://w3id.org/scischema/schema/biology/rna-seq-dge-workflow/}{\url{biology/rna-seq-dge-workflow}} & \href{https://orkg.org/templates/R1644878}{R1644878} \\
Bulk RNA-seq Library Preparation and Sequencing & \href{https://w3id.org/scischema/schema/biology/rna-seq-library-preparation/}{\url{biology/rna-seq-library-preparation}} & \href{https://orkg.org/templates/R1644668}{R1644668} \\
Fatigue Testing of Metallic Materials & \href{https://w3id.org/scischema/schema/materials/fatigue-testing/}{\url{materials/fatigue-testing}} & \href{https://orkg.org/templates/R1643108}{R1643108} \\
Steam Reforming & \href{https://w3id.org/scischema/schema/materials/steam-reforming}{\url{materials/steam-reforming}} & \href{https://orkg.org/templates/R1899386}{R1899386} \\
Solvent Casting for Polymer Composites & \href{https://w3id.org/scischema/schema/materials/solvent-casting/}{\url{materials/solvent-casting}} & \href{https://orkg.org/templates/R1900372}{R1900372} \\
Catalytic Pyrolysis of Mixed Plastic Waste & \href{https://w3id.org/scischema/schema/materials/catalytic-pyrolysis/}{\url{materials/catalytic-pyrolysis}} & \href{https://orkg.org/templates/R1642000}{R1642000} \\
Metal-Organic Cage Synthesis & \href{https://w3id.org/scischema/schema/materials/metal-organic-cage-synthesis/}{\url{materials/metal-organic-cage-synthesis}} & \href{https://orkg.org/templates/R1899097}{R1899097} \\
Dynamic Mechanical Analysis (DMA) & \href{https://w3id.org/scischema/schema/materials/dynamic-mechanical-analysis/}{\url{materials/dynamic-mechanical-analysis}} & \href{https://orkg.org/templates/R1900098}{R1900098} \\
Iterative Tomographic Image Reconstruction & \href{https://w3id.org/scischema/schema/imaging/iterative-xray-ct-reconstruction/}{\url{imaging/iterative-xray-ct-reconstruction}} & \href{https://orkg.org/templates/R1899000}{R1899000} \\
Multi-Electrode Array (MEA) Recordings & \href{https://w3id.org/scischema/schema/imaging/mea-recordings/}{\url{imaging/mea-recordings}} & \href{https://orkg.org/templates/R1898211}{R1898211} \\
Neutrino Event Reconstruction & \href{https://w3id.org/scischema/schema/physics/neutrino-event-reconstruction/}{\url{physics/neutrino-event-reconstruction}} & \href{https://orkg.org/templates/R1644988}{R1644988} \\
Stroop Task & \href{https://w3id.org/scischema/schema/psychology/stroop-task/}{\url{psychology/stroop-task}} & \href{https://orkg.org/templates/R1643165}{R1643165} \\
Torrance Tests of Creative Thinking (TTCT) & \href{https://w3id.org/scischema/schema/psychology/ttct/}{\url{psychology/ttct}} & \href{https://orkg.org/templates/R1699271}{R1699271} \\
\hline
\end{tabular}
\end{table}

\section*{Code Availability}

The \SchemaMiner{} software used for the schema-mining workflow has been described previously \cite{schema-miner,schema-miner-pro} and is openly available under the MIT License at \url{https://github.com/sciknoworg/schema-miner}. Version 3.2.5 was used in this work. The package is distributed through PyPI at \url{https://pypi.org/project/schema-miner/}; the version used for this study can be installed with \texttt{pip install schema-miner==3.2.5}. Later unpinned installations of \texttt{schema-miner} may resolve to newer releases. The software supports the staged schema-mining workflow described in the Methods section, including schema generation from process specifications, iterative processing of source papers, incorporation of expert feedback, and generation of updated schema candidates.

Custom analysis code used to generate the figures and quantitative summaries reported in this Data Descriptor is included in the Zenodo dataset under the \texttt{paper-analysis/} directory \cite{dsouza_2026_scischema}. This directory contains the scripts, input tables, and generated outputs used for the token-length analyses, schema-structure analyses, expert-rating analyses, and figure generation. No authentication is required to access the released code files. Re-running the full schema-mining workflow requires access to the relevant source publications listed in the process-level \texttt{metadata.csv} files and access to the configured LLM providers or locally available models.

\section*{Competing Interests}
The authors declare no competing interests.

% Optional: remove this section if not needed.
\section*{Acknowledgements}
%Acknowledge individuals, institutions, infrastructure, data providers, reviewers, or communities that supported the work.

For 15 of the 16 processes, the following small models were obtained from Hugging Face and run on two GPU nodes at TIB--Leibniz Information Centre for Science and Technology: Ministral-3-3B-Instruct-2512 (\url{https://huggingface.co/mistralai/Ministral-3-3B-Instruct-2512}), Ministral-3-8B-Instruct-2512 (\url{https://huggingface.co/mistralai/Ministral-3-8B-Instruct-2512}), Qwen3-4B-Instruct-2507 (\url{https://huggingface.co/Qwen/Qwen3-4B-Instruct-2507}), Ministral-3-3B-Reasoning-2512 (\url{https://huggingface.co/mistralai/Ministral-3-3B-Reasoning-2512}), Ministral-3-8B-Reasoning-2512 (\url{https://huggingface.co/mistralai/Ministral-3-8B-Reasoning-2512}), and Qwen3-4B-Thinking-2507 (\url{https://huggingface.co/Qwen/Qwen3-4B-Thinking-2507}). One node was equipped with two NVIDIA L40S GPUs and the other with two NVIDIA H100 GPUs. Two processes were run concurrently on each node, allowing up to four processes to run simultaneously. The large models---Gemma-3-27B-IT, Qwen3-30B-A3B-Instruct-2507, Qwen3-235B-A22B, Mistral-Large-3-675B-Instruct-2512, DeepSeek-R1-Distill-Llama-70B, and Qwen3-30B-A3B-Thinking-2507---were accessed through the KISSKI Chat AI API (\url{https://kisski.gwdg.de/en/leistungen/2-02-llm-service/}). The service is hosted by the Gesellschaft f\"ur wissenschaftliche Datenverarbeitung mbH G\"ottingen (GWDG) and made available to public universities and research institutes in Lower Saxony \cite{doosthosseini2026saia}.

For Catalytic Pyrolysis of Mixed Plastic Waste, the model experiments were conducted through OpenRouter (\url{https://openrouter.ai/}) and using in-house GPU infrastructure at the University of Alicante.

Within the 16-process SciSchema.org collection, the domain experts contributed to the curation of their respective scientific-process schemas as follows: A. Bossler, A. Fullana, and E. Bas, Catalytic Pyrolysis of Mixed Plastic Waste; S. Ather, Iterative Tomographic Image Reconstruction; D. Circi, A. Chen, and L. C. Brinson, Solvent Casting for Polymer Composites; A. Columbus, Polymerase Chain Reaction; G. Demetriou, CRISPR-Cas9; D. Jeong, Fatigue Testing of Metallic Materials; T. Kumar, RNA-Seq Differential Gene Expression Analysis Workflow; F. Kr\"uger, S. Genehr, and K. Budde-Sagert, Electrical Cell Stimulation; A. Leonescu, Metal-Organic Cage Synthesis; F. Lodola and C. Florindi, Multi-Electrode Array Recordings; G. Balasubramanya Murthy, Stroop Task; S. Olagbile, Steam Reforming; N. Riasat, Bulk RNA-seq Library Preparation and Sequencing; Y. Sha, Torrance Tests of Creative Thinking; K. Shen, Dynamic Mechanical Analysis; and S. Yang, Neutrino Event Reconstruction.

\section*{Funding}
%List funding sources, grant numbers, funders, and recipient authors where applicable.

A. Bossler was supported by the Spanish Ministry of Science, Innovation and Universities through the REBORN project (grant PID2021-126284OB-I00). D. Circi, A. Chen, and L. C. Brinson acknowledge support from the U.S. National Science Foundation through NSF DMREF CMMI-2323978 and NSF DGE-2022040. A. Columbus was supported by the Vivien Thomas Scholars Initiative at Johns Hopkins University. F. Kr\"uger, S. Genehr, and K. Budde-Sagert were supported by the German Research Foundation (DFG) through SFB 1270/3 ELAINE (grant 299150580). F. Lodola was supported by the Italian Ministry of Universities and Research through the PRIN 2022 project (grant 2022-NAZ-0595). All remaining authors were supported by their respective institutions.

% BibTeX bibliography. The included naturemag.bst gives a compact numbered style.
\bibliographystyle{naturemag}
\bibliography{references}

\end{document}